\documentclass[prl,preprintnumbers,preprint,superscriptaddress,showpacs]{revtex4}
\usepackage{multirow,graphicx,color}

\begin{document}
\title{Neutral and charged excitations in carbon fullerenes from
first-principles many-body theories}

\author{Murilo L. Tiago}
\affiliation{Oak Ridge National Laboratory, Oak Ridge, TN, 37831} 
\author{P. R. C. Kent}
\affiliation{Oak Ridge National Laboratory, Oak Ridge, TN, 37831}
\author{Randolph Q. Hood}
\affiliation{Lawrence Livermore National Laboratory, Livermore, California, 94550}
\author{Fernando A. Reboredo}
\affiliation{Oak Ridge National Laboratory, Oak Ridge, TN, 37831}

\date{\today}

\begin{abstract} 
We investigate the accuracy of first-principles many-body theories
at the nanoscale
by
comparing the low
energy excitations of the carbon fullerenes C$_{20}$, C$_{24}$,
C$_{50}$, C$_{60}$, C$_{70}$, and C$_{80}$ with experiment.  
Properties are calculated
via the GW-Bethe-Salpeter Equation (GW-BSE) and diffusion Quantum
Monte Carlo (QMC) methods.  We critically compare these theories and
assess their accuracy against available photoabsorption and
photoelectron spectroscopy data. The first ionization potentials are
consistently well reproduced and are similar for all the fullerenes
and methods studied.  The electron affinities and first triplet
excitation energies show substantial method and geometry dependence.
These results establish the validity of many-body
theories as viable alternative to density-functional theory in
describing electronic properties of confined carbon nanostructures. We
find a correlation between energy gap and stability of fullerenes. We
also find that the electron affinity of fullerenes is  very high and
size-independent, which explains their tendency to form compounds with
electron-donor cations.

\end{abstract}


\maketitle

\section{Introduction}
Carbon fullerenes form a
remarkable series of molecules whose physical and chemical properties
are relatively well known. Their properties can be controlled through size, shape, and
structure\cite{dresselhaus}. They can be functionalized, doped with
endohedral or exohedral atoms, or polymerized \cite{osawa}.
C$_{60}$ is the most stable and well-studied molecule in the
series. It has been 
widely
used 
{\it e.g.}
in the design of efficient organic
photovoltaic cells \cite{XueURF04,YangSF05} and as precursor in the
growth of carbon nanotubes and nanopeapods
\cite{PfeifferHPKLSK07,HernandezMSRTNTLC03}.
The strength of
carbon $sp^2$ bonds makes fullerenes extremely stable, 
and
the
ability of carbon to form bonds with different hybridization levels
makes possible the existence of 
both hexagonal and
pentagonal rings 
and
hence a rich variety of sizes and shapes\cite{dresselhaus,osawa}. In
fact, carbon is the main constituent of a wide range of nanostructures
from diamondoids to organic molecules, including fullerenes and
nanotubes. It is likely that nano-electronic devices with desired
properties can be created through intentional selection of specific
carbon nanostructures\cite{dresselhaus,RutherglenB07}. In order to
systematically design such devices it is necessary to develop a
thorough understanding of the electronic and optical properties of the
nanostructure, either isolated or in a functionalized medium. Detailed
knowledge of the electronic structure of fullerene molecules is the
key to explore several properties of those systems. As an example, the
conductance of fullerene-based junctions depends crucially on the
relative energy of electronic orbitals on the molecule with respect to
the chemical potential on the leads \cite{NeatonHL06}. 

Experimental characterization is often hindered 
at the nanoscale
and
therefore theory plays an increasingly important role in nano-science
as a guide for experimental interpretation. While it is often claimed
that theory can be used to design nanomaterials with targeted
properties, for theory to have predictive power it is crucial to
validate it in those precious cases where the comparison with
experiment can be made. Carbon compounds including nanotubes,
\cite{barone05,spataru}
adamatanes and fullerenes are among the few nanomaterials 
that have been characterized in terms of structure and composition.

In this paper we apply the two most accurate approaches for
calculating electronic/optical properties, the GW-Bethe Salpeter
equation (GW-BSE) approach \cite{onida02,aulbur00} and diffusion
quantum Monte Carlo (QMC)\cite{WMCFoulkesRMP2001}. These methods are
complementary: the GW-BSE approach is based primarily on approximate
Green's functions and in-principle yields the entire optical spectrum,
while QMC uses a stochastic sampling approach that is in-principle
exact for the lowest energy excitation of each
symmetry\cite{WMCFoulkesRMP2001}. Due to recent algorithmic advances
and a favourable scaling with system size, both of these methods can
provide the optical properties of nanostructures with one hundred
atoms or more using existing parallel computers
\cite{tiago06,WilliamsonGHPG02}.  While GW-BSE and QMC have previously
shown good agreement for small molecules
\cite{aulbur00,onida02,WilliamsonGHPG02,rohlfing00}, this is the first
comprehensive study of these methods for these much larger yet
prototypical nanostructures. Owing to the complexity involved in both
QMC and GW-BSE when applied to electronically confined systems, not
much is known about the accuracy of either theory in confined
nanostructures.

We study a range of fullerenes ( C$_{20}$, C$_{24}$, C$_{50}$,
C$_{60}$, C$_{70}$, and C$_{80}$ ), with the aim of answering these
questions: (1) Do these methods accurately predict the optical and electronic
properties, particularly for the best experimentally characterized
C$_{60}$ molecule?  (2) What are the trends in the physical properties
across the fullerene series? (3) Where are improvements in each method
required?

\section{Method}

\begin{figure}
\includegraphics[width=12cm]{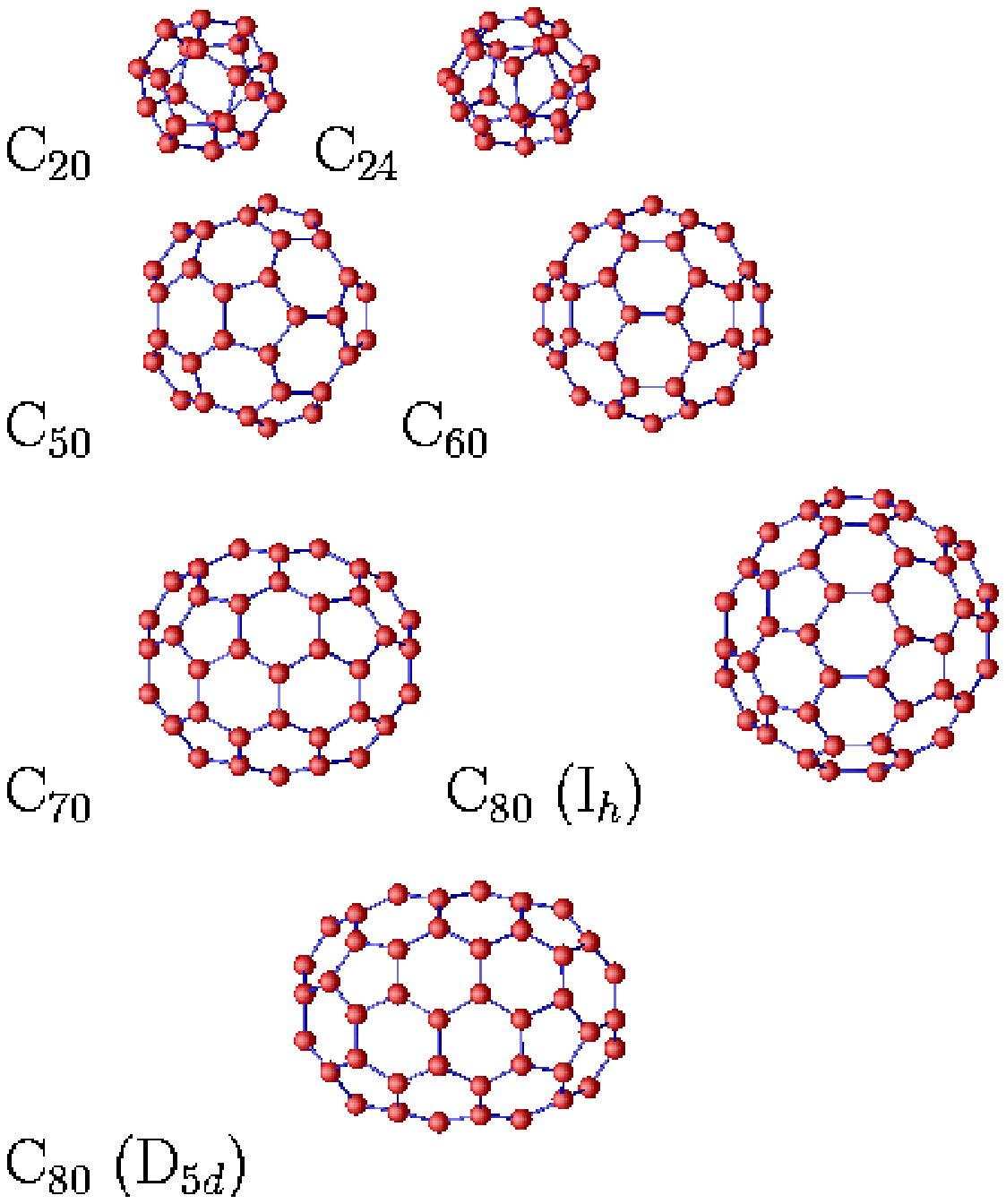} 
\caption{(Color online) Structures of the fullerenes
with corresponding point group representations in parentheses: 
C$_{20}$ (I$_{h}$), C$_{24}$
  (D$_{6d}$), C$_{50}$ (D$_{5h}$), C$_{60}$ (I$_h$), C$_{70}$ (D$_{5h}$),
  C$_{80}$ (I$_{h}$), C$_{80}$ (D$_{5d}$).}
\label{f-balls}
\end{figure}

Fullerenes are routinely produced in a wide range of
sizes, from C$_{20}$ to C$_{100}$ or larger, and many
isomers have been identified \cite{zimmerman91}. The existence of
several low-energy isomers for each molecular size poses a difficulty
in modeling fullerenes. For each computed size we have chosen
high-symmetry structures that satisfy the isolated pentagon rule
\cite{dresselhaus} as closely as possible. Figure \ref{f-balls}
depicts the isomers investigated here, with the corresponding point
groups.  The atomic coordinates were obtained by minimizing the total
energy within norm-conserving pseudopotential calculations using DFT
and the Perdew-Burke-Ernzerhof (PBE) functional
\cite{JPPerdewPRL1996}.  Choices of grid point separation, energy
cutoffs and supercell sizes were such that Kohn-Sham eigenvalues 
of bound orbitals
are
converged to 0.02~eV or better and bond lengths to 1\% or
better.
Atom coordinates are included in the supporting information
accompanying this paper\cite{EPAPS}.

Owing to its simplicity, density-functional theory (DFT) 
has been 
used
to predict the electronic properties of
nanostructures. Nevertheless, it has two major deficiencies:
(1) It can only 
describe
the energy of 
optical
excitations
through {\it ad hoc} approaches such as constrained DFT
\cite{martin};
and (2) 
it is not possible to
systematically 
improve the description of exchange and correlation
effects within DFT because the exact exchange-correlation functional
cannot be calculated in closed form \cite{martin}. In contrast,
many-body approaches can be systematically improved, at the expense of
increased numerical complexity.

In the framework of the GW approximation \cite{aulbur00},
electron self-energies are calculated in terms of the Green's function
$G$ and the screened Coulomb interaction $W$ as:
\begin{equation}
\Sigma(1,2) = i G(1,2) W_0 (1,2) \; \;,
\label{e-gw0}
\end{equation}
where screening is commonly evaluated in the random-phase
approximation (RPA).  We can modify Eq. (\ref{e-gw0}) by replacing the
RPA with time-dependent DFT with an adiabatic, local-density kernel
(TDLDA). In order to retain Feynman diagrams with the same order, we
add a properly symmetrized vertex term together with the TDLDA
screened Coulomb interaction \cite{tiago06}:

\begin{equation}
\Sigma(1,2) = i \int {\rm d}(3) G(1,3) W(3,2) \Gamma (1,2;3) +
      \mbox{symm.} \; \; .
\label{e-gwf}
\end{equation}
At its lowest level of approximation, the Green's function and the
screened Coulomb interaction above are calculated within DFT
\cite{aulbur00}.  Further improvement in the theory is attained by
imposing self-consistency among the self-energy $\Sigma$, Green's
function $G$ and screened Coulomb interaction $W$ \cite{aulbur00}
We
impose self-consistency in an iterative scheme. In the first
iteration, we calculate the self-energy using DFT electronic orbitals
and Eq. (\ref{e-gwf}). In subsequent iterations, we use a scissors
operator to correct the energy of DFT orbitals, according to the
self-energy calculated in the previous iteration. 
We stress the fact that the scissors operator is constructed using
information from the previous iteration only. No free parameters are involved.
We stop when
electron affinity and ionization potential 
are converged to within
0.02 eV. Usually, this is reached in less than five iterations. Tests
performed on smaller molecules 
( Cl$_2$, benzene, CO, SiH$_4$)
indicate that 
going behond the scissors operator method
would improve the final
results by 0.1~eV or less.
Scissors operators have been used
extensively in the context of electronic band structure of bulk
semiconductors (see Ref. \onlinecite{aulbur00} and references therein).
In the following, we refer to Eqs.
(\ref{e-gw0}) and (\ref{e-gwf}) as GW$_0$ and GW$_f$ approximations
respectively. We denote by scGW$_f$ approximation the self-consistent
solution of Eq. (\ref{e-gwf}),
where self-consistency is imposed through a scissors operator. 
After the quasi-particle
orbitals are determined, we calculate optical excitations by
diagonalizing the Bethe-Salpeter equation for electrons and holes
following standard methodologies \cite{onida02,rohlfing00,tiago06}.
The BSE method has been used successfully to predict excitation
energies of organic molecules \cite{tiago06}, CdSe nanocrystals
\cite{delpuerto06} and hydrogenated silicon clusters
\cite{rohlfing00}.

In the QMC methodology (for a review, see Ref.
\cite{WMCFoulkesRMP2001}), excitation energies are obtained via the
difference in total energy between the ground state and individually
constructed excited states. We use the diffusion Monte Carlo (DMC)
technique with trial wave-functions consisting of a single Slater
determinant of single-particle orbitals multiplied by a Jastrow
correlation function \cite{casino_07}. The most significant
approximation is the fixed-node approximation, whereby the Fermion
nodes of a trial wave-function are used in place of the exact nodes,
introducing a variational error. Although DMC calculations for large
systems additionally require the use of pseudopotentials, for
carbon-based systems this approximation is secondary. We evaluate the
pseudopotentials within the locality approximation. A test calculation
for C$_{20}$ using a variational pseudopotential
evaluation\cite{MCasulaPRB2006} found a global shift of $0.8 \pm 0.1 $
eV but identical energy differences to within $\pm 0.1$ eV,
demonstrating that this error is small in the fullerenes. In practice,
the accuracy of the results obtained for excitations is solely
determined by the accuracy of the ground and excited state nodal
surfaces. In contrast, the GW calculations are dependent on the
detailed shape of the Kohn-Sham orbitals and the initial Kohn-Sham
eigenvalues.  Given exact nodes for the ground state and for the first
excited state of a given symmetry, DMC yields in-principle exact
results\cite{WMCFoulkesRMP2001}. 
In general, there are additional symmetry related qualifications  
\cite{FoulkesHN99}.
In our calculations we use the
Kohn-Sham orbitals from the ground state DFT calculation in the Slater
determinant to effectively define the nodal surface of the trial
wavefunction. DMC was used recently to calculate excitation energies
and ionization energies of diamondoids yielding very good results
\cite{drummond05}.

To make comparisons between GW-BSE and QMC on an equal footing we used
the same norm conserving pseudopotential in each set of
calculations. The initial QMC wave-functions were obtained from a
local-density approximation (LDA) DFT calculation, while for GW
calculations we additionally investigated the
PBE\cite{JPPerdewPRL1996} gradient corrected functionals.  GW$_f$ electron
affinities and vertical ionization potentials obtained using the LDA
are higher than the ones obtained using PBE by a fraction of
electron-volt: 0.1~eV in C$_{60}$ and 0.3~eV in C$_{20}$. The impact
of replacing the LDA with PBE on Kohn-Sham eigenvalues is even
smaller: less than 0.05~eV. In view of these small differences we
concentrate on LDA-derived results.

\section{Results}

We have investigated two isomers of C$_{80}$.  DFT calculations have
already indicated that C$_{80}$ exists in a large number of isomers,
where the oval shaped structure (point group D$_{\rm 5d}$) is the most
stable one \cite{furche01,bauernschmitt98}. Our DFT calculations
predict that the icosahedral isomer (I$_h$) is more energetic than the
D$_{\rm 5d}$ isomer by 0.8~eV, confirming earlier calculations
\cite{furche01}. This difference increases to $1.74\pm 0.10$ eV in
QMC.

Table \ref{t-ip} shows the vertical ionization
potential (IP) calculated within the GW and QMC theories. The
ionization potential was also calculated using DFT only, as the
difference in total energy between the neutral and the charged
molecules (``$\Delta$SCF-DFT'') column. The IP is approximately 7.5~eV for
all of the theories, with no well defined trend in IP with size
apart from a slow decrease with increasing number of atoms.
The analyzed structures have different curvatures, caused by the
different positions of pentagonal rings from molecule to
molecule. Therefore, geometry and symmetry become important factors in
determining the ionization potential. A classical electrostatic model
based simply on a charged hollow sphere can only predict qualitatively
the ionization potential of fullerenes \cite{Brus83,SeifertVS96}.
Except for C$_{60}$ and
C$_{70}$, the QMC ionization potential is lower than the GW$_f$
ionization potential. The discrepancy between the two approaches is
around 0.2~eV, close to the numerical accuracy associated with each
theory
\footnote{ QMC statistical errors are 0.1~eV to 0.2~eV. All GW results have
estimated systematic convergence errors of less than 0.2~eV.  }.
Except for GW$_0$ all the methods predict IPs close to experiment.
Self-consistency in the GW method is found to give very small
improvement on the already accurate IP.
Table \ref{t-ip} also shows that the GW$_f$ approximation predicts an IP
much more accurate than the GW$_0$ approximation. Therefore, including
vertex corrections and polarizability effects beyond RPA are an
improvement on the theory, confirming similar observations in benzene
and naphthalene molecules \cite{tiago06}. 

\begin{table}
\caption{Vertical ionization potentials calculated from DFT, QMC and GW
theories. Error bars in QMC values are indicated between
parentheses.
All energies in eV.}
\label{t-ip}
\begin{ruledtabular}
\begin{tabular}{ccccccc}
 & $\Delta$SCF-DFT & QMC & GW$_0$ & GW$_f$ & scGW$_f$ & Exp. \\ \hline
C$_{20}$ & 7.31 & 7.27(11) & 7.99 & 7.35 & 7.41 & \\
C$_{24}$ & 7.77 & 7.70(10) & 8.49 & 7.86 & 7.81 & \\
C$_{50}$ & 7.29 & 7.29(14) & 7.97 & 7.33 & 7.35 & 7.61$^a$ \\
C$_{60}$ & 7.61 & 7.86(21) & 8.22 & 7.70 & 7.86 & 7.6$^b$ \\
C$_{70}$ & 7.54 & 7.69(12) & 8.12 & 7.53 & 7.45 & 7.47$^c$ \\
C$_{80}$ (D$_{5d}) $ & 6.67 & 6.30(10) & 7.24 & 6.59 & 6.65 & 6.84$^a$ \\
C$_{80}$ (I$_h$) & 6.86 & 6.91(10) & 7.45 & 6.90 & 6.95 & \\ \hline
average error & -0.10 & -0.09 & 0.51 & -0.09 & -0.05 \\
root mean square error & 0.18 & 0.36 & 0.52 & 0.20 & 0.21 \\
\end{tabular}
\end{ruledtabular}
$^a$ Ref. \onlinecite{zimmerman91}.
$^b$ Ref. \onlinecite{lichtenberger91}
$^c$ Ref. \onlinecite{lichtenberger92}
\end{table}

Ultra-violet photoelectron spectroscopy (UPS) can determine not only
the first ionization potential but also higher ionization energies,
thus giving direct access to the electronic structure of the
material. Lichtenberger and collaborators \cite{lichtenberger91} have
identified five ionization bands in the C$_{60}$ spectrum. Table
\ref{t-ups} shows the assignment of the ionization bands and
comparison between measured and calculated vertical ionization
energies.  Higher-order ionization energies can only be calculated
within DFT if a minimization constraint is used. Calculation results
presented on column ``CDFT'' on Table \ref{t-ups} were obtained by
minimizing the total energy of the cation C$_{60}^+$ with the
constraint that the ejected electron occupied a specific molecular
orbital (HOMO, HOMO-1, or deeper orbitals). Whereas constrained DFT
predicts very accurately the first two ionization energies, it
increasingly underestimates the higher ionization energies. The
discrepancy seems to be more than 1~eV in the ionization band at 12.4 to
13.8~eV. In contrast, the self-consistent GW$_f$ approximation
predicts ionization energies no more than 0.35~eV away from the
experimental data
over the same energy range.

\begin{table}
\caption{Orbital assignment and vertical ionization energy for the
  highest occupied molecular orbitals in C$_{60}$. Column ``C-DFT''
  shows constrained DFT results. Columns GW$_0$ and
  GW$_f$ are non-self-consistent GW calculations, with and without
  vertex corrections respectivelly. Column scGW$_f$ includes
  self-consistency and vertex corrections. Experimental data quoted
  from Ref. \onlinecite{lichtenberger91}. Experimental data spanning
  multiple rows indicate ionization bands that could not be
  resolved. All energies in eV.}
\label{t-ups}
\begin{ruledtabular}
\begin{tabular}{cccccc}
 & C-DFT  & GW$_0$ & GW$_f$ & scGW$_f$ & Exp. \\ \hline
h$_u$ & 7.61 & 8.22 & 7.70 & 7.86 & 7.6 \\ \hline
g$_g$ & 8.78 & 9.33 & 8.85 & 9.23 & \multirow{2}{*}{8.95} \\
h$_g$ & 8.90 & 9.42 & 8.93 & 9.25 &  \\ \hline
h$_u$ & 10.47 & 11.93 & 11.40 & 11.93 &\multirow{3}{*}{10.82-11.59} \\
g$_u$ & 10.50 & 11.00 & 10.51 & 11.07 & \\
t$_{2u}$ & 11.03 & 11.46 & 11.02 & 11.6 & \\ \hline
h$_g$ & 10.79 & 12.19 & 11.65 & 12.2 & \multirow{4}{*}{12.43-13.82} \\
g$_u$ & 11.79 & 13.23 & 12.73 & 13.36 & \\
t$_{1g}$ & 12.28 & 13.60 & 13.11 & 13.76 & \\
h$_g$ & 12.33 & 12.74 & 12.32 & 13.04 & \\
\hline
\end{tabular}
\end{ruledtabular}
\end{table}

The electron affinity (EA), shown in Table \ref{t-ea} is also
predicted differently by QMC and the various levels of GW theory. The
former theory underestimates the electron affinity by as much as 1~eV
relative to the latter one. Inclusion of vertex corrections (from
GW$_0$ to GW$_f$) reduces the electron affinity by approximately
0.6~eV. Inclusion of self-consistency (from GW$_f$ to scGW$_f$)
increases the self-energy by a further 0.5~eV, bringing the calculated
results in very good agreement with experimental data and with
$\Delta$SCF-DFT predictions. The electron affinity is found to be
around 2 to 4~eV. All theories employed in this work predict that
C$_{20}$ has the smallest electron affinity whereas C$_{80}$(I$_h$)
has the largest electron affinity. C$_{60}$ is well known to
be easily ionizable \cite{osawa,dresselhaus,WangDW99,WangWHKW06}. The
present results also show that less stable fullerenes are easily
ionizable as well, which opens the path to functionalize fullerenes
of different sizes.

\begin{table}
\caption{Electron affinity, or vertical detachment energy, calculated
from DFT, QMC and GW theories. Error bars in QMC values are indicated
between parentheses.  All energies in eV.}
\label{t-ea}
\begin{ruledtabular}
\begin{tabular}{ccccccc}
 & $\Delta$SCF-DFT & QMC & GW$_0$ & GW$_f$ & scGW$_f$ & Exp. \\ \hline
C$_{20}$ & 2.17 & 1.76(11) & 3.55 & 2.92 & 2.36 & 2.25$^a$ \\
C$_{24}$ & 3.04 & 2.57(11) & 4.19 & 3.55 & 2.88 & \\
C$_{50}$ & 3.73 & 3.52(14) & 4.75 & 4.12 & 3.73 & $<$ 3.10$^b$ \\
C$_{60}$ & 2.94 & 2.23(19) & 3.87 & 3.33 & 2.98 & 2.69$^c$ \\
C$_{70}$ & 2.96 & 2.46(11) & 3.98 & 3.35 & 2.83 & 2.76$^d$ \\
C$_{80}$ (D$_{5d}) $ & 3.46 & 3.25(10) & 4.62 & 3.91 & 3.88 &  3.70$^b$ \\
C$_{80}$ (I$_h$) & 3.98 & 3.90(11) & 5.17 & 4.61 & 4.38 & \\ \hline
average error & 0.03 & -0.42 & 1.25 & 0.53 & 0.16 \\
root mean square error & 0.20 & 0.43 & 1.16 & 0.56 & 0.18 \\
\end{tabular}
\end{ruledtabular}
$^a$ Ref. \onlinecite{prinzbach06}.
$^b$ Ref. \onlinecite{yang87}.
$^c$ Ref. \onlinecite{WangDW99}.
$^d$ Ref. \onlinecite{WangWHKW06}.
\end{table}

One important result of this work is that GW$_0$ and GW$_f$ are
shown to be 
consistent in the ``electronic gap'', defined as the
difference between ionization potential and electron affinity. This
definition of electronic gap is the closest analogue of band gap to
isolated molecules \cite{aulbur00}. Although GW$_0$ and GW$_f$ predict
different values for IP and EA, the differences cancel out in the
electronic gap. The reason for this cancellation is that contributions
arising from LDA screening, such as the vertex $\Gamma$, have similar
magnitude when evaluated at the highest occupied molecular orbital
(HOMO) and at the lowest unoccupied molecular orbital (LUMO). As a
consequence of this cancellation, the contributions from LDA screening
are often neglected in calculations of electronic gap in crystals
\cite{aulbur00}. In molecular systems, these contributions are not
negligible because they affect IP and EA separately
\cite{tiago06,morris07}. Self-consistency has been found to
systematically increase the electronic gap compared to
non-self-consistent GW methodologies in a variety of different
materials\cite{aulbur00,Kotaniv07}. Our calculations confirm the same
behavior in finite systems, and they also indicate that
self-consistency affects the energy of unoccupied orbitals more than
of occupied orbitals. Self-consistency could be more important in
fullerenes than in most small molecules studied so far
\cite{rohlfing00,tiago06,delpuerto06,ismail-beigi03} because of the
narrower HOMO-LUMO gap.

There is an ongoing debate in the literature concerning the importance
of vertex corrections \cite{aulbur00,tiago06,morris07}. The GW$_f$
approximation has vertex corrections included in two ways: explicitly
through the function $\Gamma$ in Eq. (\ref{e-gwf}), and implicitly
through the screened Coulomb interaction (see {\it e.g.}  Eq. 17 and
18 in Ref. \cite{tiago06}). We have observed that most of the
difference between GW$_f$ and GW$_0$ calculation results originates
from the explicit contribution rather than the implicit
contribution. Matrix elements of the ``explicit vertex contribution'',
defined as $\Delta \Sigma = i (GW\Gamma - GW)$ have approximately the
same magnitude at the HOMO and the LUMO of each molecule. As a
function of molecule size, $\Delta \Sigma$ fluctuates around 0.6~eV to
0.7~eV across the family of fullerenes studied, with the exception of
C$_{20}$ where it is around 0.8~eV. The ``implicit vertex
contribution'', $i(GW - GW_0)$ is typically five times smaller in
magnitude. Self-consistency following the prescription presented here
does not modify significantly the strength of vertex
contributions. This behavior is similar to the one observed in small
oligoacenes (benzene and naphthalene) \cite{tiago06} but it is
somewhat different from the scenario observed in silicon
nanocrystals. There, the explicit vertex contribution is no more than
around 0.3~eV and cancellation between explicit and implicit contributions is
more complete.

Table \ref{t-triplet} shows the calculated and measured excitation
energy of the first spin-triplet electronic states in each fullerene.
Experimental data were obtained from phosphorescence decay
measurements, and they include a Stokes shift absent in the
theoretical calculations. Based on $\Delta$SCF-DFT calculations, we
estimate the Stokes shift to be no more than 0.2~eV. Independent
evidence for small Stokes shift is found in the photoelectron spectra
of C$_{60}^-$ and C$_{70}^-$, which show a high 0-0 line followed by
lower vibrational side bands at higher
energy\cite{WangDW99,WangWHKW06}.  TDLDA gives good excitation
energies, with contrasts with the well-known gap underestimation of
carbon nanotubes \cite{BachiloSKHSW02} and of periodic systems in
general \cite{onida02}.  Self-consistent GW-BSE is also very accurate,
having a discrepancy from experimental data comparable to the Stokes
shift.  The QMC data is too high by approximately 0.8~eV, and the
non-self-consistent GW-BSE data is too low by approximately 0.5~eV.

\begin{table}
\caption{Excitation energy of the first spin-triplet state. Calculated
values under the ``GW$_f$-BSE'' column were obtained using the GW$_f$
approximation in the electron self-energy.
Column ``scGW$_f$-BSE'' indicates results obtained with the self-consistent GW$_f$
approximation.
Error bar in QMC values
are indicated between parentheses.  All energies in eV.}
\label{t-triplet}
\begin{ruledtabular}
\begin{tabular}{cccccc}
 & TDLDA & QMC & GW$_f$-BSE & scGW$_f$-BSE & Exp. \\ \hline
C$_{20}$ & 0.57 & 0.87 (8) & -0.12 & 0.62 & \\
C$_{24}$ & 0.43 & 0.90(14) & -0.07 & 0.67 & \\
C$_{50}$ & 0.20 & 0.61(13) & -0.17 & 0.24 & \\
C$_{60}$ & 1.52 & 2.34(19) & 1.14 & 1.65 & 1.58$^a$ \\
C$_{70}$ & 1.62 & 2.31(12) & 1.01 & 1.45 & 1.55$^b$ \\
C$_{80}$ (D$_{5d}) $ & 0.41 & 0.07(10) & -0.15 & 0.20 & \\
C$_{80}$ (I$_h$) & 0.11 & 0.26(10) & -0.35 & -0.06 & \\ \hline
average error & 0.01 & 0.76 & -0.49 & -0.02 \\
root mean square error & 0.05 & 0.76 & 0.49 & 0.09 \\
\end{tabular}
\end{ruledtabular}
$^a$ Ref. \onlinecite{giuffreda01}.
$^b$ Ref. \onlinecite{warntjes96}.
\end{table}

The difference in the QMC results must result from a poor cancellation
of nodal error between ground and triplet excited states. For small
molecules it is possible to improve the nodal surface by constructing
multideterminant trial wavefunctions or performing additional
wavefunction optimization\cite{UmrigarTFSH07}, but these techniques
are not yet practical for the large fullerenes.
Clearly fullerenes represent a good benchmark for methods to improve
the nodal surfaces.

The GW-BSE results reflect the discrepancies in the electron affinity:
since low-energy optical excitations usually involve the LUMO,
excitation energies obtained without self-consistency are
underestimated by approximately 0.6~eV (column ``GW$_f$-BSE'' in Table
\ref{t-c60}). We further examine the observed discrepancies by
computing higher-energy neutral excitations of C$_{60}$. Table
\ref{t-c60} presents excitation energies of the first few spin-singlet
states and the projection of each excited state on the subspace of
electron transitions from the HOMO to the LUMO. This projection is
simply a sum over overlap integrals between the GW-BSE eigenstates $|S
\rangle$ and the many-body states obtained by promoting one electron
from the HOMO to the LUMO in the ground state $|G \rangle$:

\begin{equation}
P_{H-L} = \sum_j^{\mbox{HOMO} } \sum_i^{\mbox{LUMO} } | \langle S
| a^\dagger_i a_j | G \rangle |^2 \;\; ,
\label{e-projection}
\end{equation}
where $a^\dagger_i$ and $a_j$ are many-body creation and destruction
operators respectively. This projection is not readily available in
DMC since only the probability density consisting of the product of
the QMC ground state and DFT-based trial wavefunction is available.

Identification of the various excitations in C$_{60}$ is 
facilitated  because the molecule is highly symmetric, leading to high degeneracy
and wide energy separation between states. The HOMO and the LUMO of
C$_{60}$ have symmetry representation h$_u$ and t$^1_{u}$
respectively. Electronic transitions between them give rise to four
multiplets
T$^1_g$, H$_g$, T$^2_g$, G$_g$. Table \ref{t-c60} shows that
the excitation energy of all four 
multiplets
is underestimated within the
non-self-consistent GW$_f$ approximation, typically by 0.6~eV. This is
a direct consequence of the severe overestimation of the electron
affinity if self-consistency is not imposed. Some of the higher-energy
excitations do not involve the LUMO, such as excitation T$^2_u$.  For
them, the GW-BSE prediction is compatible with experimental data.

\begin{table}
\caption{First spin-singlet excitation energies in C$_{60}$. All
energies in eV.
We use the same notation of Table \ref{t-triplet}.
 Column labeled ``H-L'' indicates the projection of
this excitation onto the HOMO-LUMO subspace (see text). Experimental
data from Ref. \onlinecite{leach92}.}
\label{t-c60}
\begin{ruledtabular}
\begin{tabular}{cccccc}
 & TDLDA & GW$_f$-BSE & scGW$_f$-BSE & H-L & Exp. \\ \hline
T$^1_g$ & 1.69 & 1.47 & 1.88 & 99 \% & 1.82-1.97
 \\
T$^2_g$ & 1.71 & 1.53 & 1.92 & 99 \% & 1.82-1.97
\\
G$_g$ & 1.64 & 1.41 & 1.82 & 99 \% &  2.00-2.21
\\
H$_g$ & 1.79 & 1.70 & 2.15 & 94 \% &  2.25-2.32
\\ \hline
\end{tabular}
\end{ruledtabular}
\end{table}

Finally, Table \ref{t-triplet} shows that the most stable fullerenes,
namely C$_{60}$ (I$_h$) and C$_{70}$ (D$_{5h}$), have the largest
triplet excitation energies. Less stable fullerenes have small triplet
excitation energy, of the order of 0.4~eV, whereas C$_{80}$(I$_h$) is
predicted to have the smallest triplet energy: 0.11~eV
(TDLDA). Knowing that triplet excitation energies are related to the
energy gap, this trend shows perfectly correlation between stability
and energy gap: stable fullerenes have wide gap between HOMO and LUMO,
unstable fullerenes have narrow gap between HOMO and LUMO.

\section{Conclusion}

We have calculated and analyzed the low energy
excitations of fullerenes C$_{20}$ to C$_{80}$ using three theoretical
methods: density functional theory, many-body GW-BSE and diffusion
Quantum Monte Carlo. Overall these methods give an accurate
description of the optical properties. Ionization potentials are
approximately constant and well predicted by all methods. The more
stable fullerenes (C$_{60}$ and C$_{70}$) are found to have highest
excitation energy. Stability correlates well with energy gap. We find
a weak dependence of electron affinity with molecule size, with
important implications for doping and functionalization of
fullerenes. We find that two ingredients are essential to bring GW-BSE
predictions to agree with experimental data: vertex corrections, which
are responsible for an almost rigid shift of electronic orbitals with
respect to the vacuum level, and self-consistency, 
which widens the energy gap between occupied orbitals 
and unoccupied orbitals. 
 QMC
calculations of the lowest energy triplet excitation using a single
determinant of Kohn-Sham orbitals give results higher than experiment
by approximately 0.8~eV, indicating 
that the DFT-derived trial wave-functions presumably 
give nodal errors of that magnitude. Therefore
while QMC calculation of even larger systems are possible as the systems size increases so
does the nodal error. Methods to improve the nodal structure 
are required to achieve a {\it total} energy resolution of 
0.1~eV. 
We believe the above observations will guide method selection for
excitations in nanostructures. Clearly, the robust design of optical
nanoarchitectures will require further developments in many-body
theories such as GW-BSE and QMC. We also encourage more precise
experimental determination of  electronic excitations in 
well-characterized
nanostructures to provide a more severe test to theory.

Research performed at the Materials Science and Technology Division,
sponsored by the Division of Materials Sciences Engineering BES,
U.S. DOE, and at the Center for Nanophase Materials Sciences,
sponsored by the Division of Scientific User Facilities,
U.S. Department of Energy under contract with UT-Battelle, LLC.
Work at the Lawrence Livermore National Laboratory was performed under the auspices of the U.S. Department of Energy 
under Contract DE-AC52-07NA27344.
Computational support was provided by the Texas Advanced Computing
Center (TACC) and the National Energy Research Scientific Computing
Center (NERSC) and Lawrence Livermore National Laboratory.


\end{document}